\documentclass[journal,onecolumn,draftcls]{IEEEtran}

\newcommand{\figwidth}{\columnwidth}

\usepackage{cite}
\usepackage{amssymb}

\ifCLASSINFOpdf
  \usepackage{graphicx}
\else
\fi

\usepackage[cmex10]{amsmath}

\begin{document}
\title{A GOST-like Blind Signature Scheme Based on Elliptic Curve Discrete Logarithm Problem}

\author{ Hossein~Hosseini, Behnam Bahrak* and Farzad~Hessar**\\ 
				*Electrical Engineering Department, Virginia Tech University\\ 
**Electrical Engineering Department, University of Washington\\
        h\_hosseini@alum.sharif.edu, *bahrak@vt.edu, **farzad@u.washington.edu
				\thanks{Corresponding Author: h\_hosseini@alum.sharif.edu}
}

\maketitle

\begin{abstract}
In this paper, we propose a blind signature scheme and three practical educed schemes based on elliptic curve discrete logarithm problem. The proposed schemes impart the GOST signature structure and utilize the inherent advantage of elliptic curve cryptosystems in terms of smaller key size and lower computational overhead to its counterpart public key cryptosystems such as RSA and ElGamal. The proposed schemes are proved to be secure and have less time complexity in comparison with the existing schemes.
\end{abstract}


\IEEEpeerreviewmaketitle

\section{Introduction}
Blind signature is a form of digital signature in which the message is blinded before it is signed, in order to allow the requester to get a signature without giving the signer any information about the actual message or the resulting signature. Blind signatures are used to build practical offline and online untraceable electronic cash schemes \cite{Chaum90, Ferguson94, Frankel96, Radu97} and widely employed in privacy-related cryptographic protocols, such as electronic election systems \cite{Chaum84}. The paper analogy to the blind signature is  enclosing a ballot in a carbon paper lined envelope; In this way, the signer does not view the message content, and also everyone can later check the validity of the signature.

Several blind signature schemes are proposed in the literature. The first scheme, proposed by Chaum \cite{Chaum82}, was based on RSA signature. In \cite{Okamoto}, Okamoto proposed the blind Schnorr signature and Pointcheval et al. proved its security in \cite{Pointcheval}. In 1995, Camenisch et al. proposed a blind signature scheme based on the Discrete Logarithm Problem (DLP) \cite{Camenisch} and later, in 2005, Wu et al. proved its untraceability \cite{Ting}. Pointcheval developed a blinding scheme for Okamoto's signature in \cite{Pointcheval2}. In \cite{Huang}, Huang et al. presented a blind signature scheme based on GOST signature, which is the Russia's digital signature algorithm \cite{Michels}. In \cite{Nikooghadam}, an efficient blind signature scheme is presented based on the elliptic curve discrete logarithm problem.

In this paper, we propose a GOST-like blind signature scheme and three efficient educed schemes based on elliptic curve discrete logarithm problem. The schemes utilize the inherent advantage of elliptic curve cryptosystems in terms of smaller key size and lower computational overhead compared to its counterpart public key cryptosystems such as RSA and ElGamal. The schemes are proved to be correct and secure. They can be used in various cryptographic protocols where the anonymity of the requester is required.

The remainder of this paper is organized as follows. In Section 2, basic concepts of elliptic curves are presented. The GOST digital signature scheme is described in Section 3. In Section 4, the generalized scheme and three other educed schemes are elaborated and the security and performances are discussed. Finally, Section 5 concludes the paper.

\section{Elliptic Curves over Finite Fields}
The elliptic curve analogues of DLP-based schemes was independently proposed by Koblitz \cite{Koblitz} and Miller \cite{Miller}, in 1985. Since then, several cryptosystems are developed based on elliptic curve computations.

A non-super singular elliptic curve $E$ over a finite field $F_q$ is as follows:
\begin{align}
E: y^2=x^3+ax+b \mod q
\label{eq:first}
\end{align}
where $4a^3+27b \mod q\not=0$. The point $\textbf{P}=(x,y)$, where $(x,y)\in F_q\times F_q$ satisfy Equation \ref{eq:first}, together with a point at infinity, denoted by $\textbf{O}$, form an abelian group $(E,+,\textbf{O})$ whose identity element is $\textbf{O}$. 

The negative of $\textbf{P}=(x_p,y_p)$ is $-\textbf{P}=(x_p, -y_p)$. Let $\textbf{P}=(x_p,y_p)$ and $\textbf{Q}=(x_q,y_q)$ be two distinct points on an elliptic curve such that $\textbf{P}\not =-\textbf{Q}$. Then $\textbf{P}+\textbf{Q}=(x_r,y_r)$, where:
\begin{align}
x_r &=(s^2-x_p-x_q )  \mod q   \nonumber \\
y_r &= (-y_p+s(x_p-x_r ))  \mod q
\end{align}
where $s=\frac{y_p-y_q}{x_p-x_q} \mod q$. 

Doubling a point $\textbf{P}$, in case of $y_p\not =0$, results in $2\textbf{P}=(x_r,y_r)$, where:
\begin{align}
x_r &=(s^2-2x_p )  \mod q \nonumber \\
y_r &= (-y_p+s(x_p-x_r ))  \mod q
\end{align}
where $s= \frac{3x_p^2+a}{2y_p}  \mod q$.

\textbf{Definition:} Let $E$ be an elliptic curve over a finite field $F_q$ and let $\textbf{P}\in E(F_q)$ be a point of order $n$. Given another point $\textbf{Q}\in E(F_q)$, the Elliptic Curve Discrete Logarithm Problem (ECDLP) is to find the integer $d\in [0,n-1]$, such that $\textbf{Q}=d\textbf{P}$ \cite{Hankerson}.

\section{The GOST Signature Scheme}
In this section, we describe the GOST digital signature scheme \cite{Michels}.

Let $p$ and $q$ be large primes that satisfy $q|p-1$, and $g$ be an element in $Z_p^*$ with order $q$. Let $H:\{0,1\}^*\rightarrow Z_q$ be a secure hash function. The signer's secret and public key pair is $(x,y)$, where $x\in Z_q$ and $y=g^x \mod p$. Let $m$ be the message to be signed.

\textbf{Signing:} The signer chooses random number $k\in Z_q$ and computes:
\begin{align}
r &= (g^k  \mod p)  \mod q \nonumber \\
s &= xr+kH(m)  \mod q
\end{align}
The signature on message $m$ is $(r,s)$.

\textbf{Verification:} The verifier computes:
\begin{align}
v   &=H(m)^{q-2} \mod q \nonumber \\
z_1 &=sv \mod q 			\nonumber \\
z_2 &=(q-r)v \mod q		\nonumber \\
u   &=\left(g^{z_1} y^{z_2}  \mod p\right)  \mod q
\end{align}
and checks whether $u=r$.

\section{The Proposed GOST-like Blind Signature Scheme}
In \cite{Huang}, a blind signature scheme based on the GOST signature is presented. Here, we propose a GOST-like blind signature scheme based on ECDLP.

There are two participants in a blind signature scheme: a signer and a group of requesters. Initially, the signer publishes the necessary information. Then, the user sends a blinded version of the message to the signer. The signer signs the blinded message, and sends the result back to the user. Afterwards, the user extracts the signature. At the end, the validity of the signature is verified. The details of these phases are described below.

\textbf{Initialization:} First, the curve parameters must be agreed upon by signer and requester. Let $E$ be the used elliptic curve over $F_q$ and suppose that the number of $F_q$-rational points on $E$ is divisible by a sufficiently large prime $n>2^{160}$. Let $\textbf{G}$ be a point on $E$ of order $n$. Signer must have a key pair suitable for elliptic curve cryptography, consisting of a private key $d$ (a randomly selected number in the interval $[1,n-1]$) and a public key $\textbf{Q}$ where $\textbf{Q}=d\textbf{G}$.

Then the signer chooses random number $k$ in the interval $[1,n-1]$, computes $\textbf{R}=k\textbf{G}=(x_r,y_r)$ and sends $\textbf{R}$ to the requester.

\textbf{Requesting:} The requester chooses random numbers $t_1$, $t_2$ and $t_3$ in the interval $[1,n-1]$ and computes:
\begin{align}
\textbf{X}   &=(t_1 \textbf{R}+t_2 \textbf{G}+t_3 \textbf{Q})=(t_1 k+t_2+t_3 d)\textbf{G} \\
m' &=x_r t_1 (m^{-1}+t_3 )^{-1}
\end{align}
then sends $m'$ to the signer. $m'$ is an encrypted version of the message, i.e. the blinded message.

\textbf{Signing:} Signer computes the signature of the blinded message as:
\begin{align}
s'=dx_r+km'
\end{align}
and sends the result back to the requester.

\textbf{Extraction:} Requester extracts the signature of the message from the signature of the blinded message, by computing:
\begin{align}
s=m(t_1 s'm'^{-1}+t_2 )
\end{align}
and declares the pair $(\textbf{X},s)$ as the signature on $m$.

\textbf{Verification:} The legitimacy of the signature $(\textbf{X},s)$ for the message $m$ is verified by examining:
\begin{align}
s\textbf{G} = m\textbf{X} + \textbf{Q}.
\end{align}
The various phases of the proposed scheme are summarized in Figure 1. 

\begin{figure}[t]%
\centering
\includegraphics[width=\figwidth]{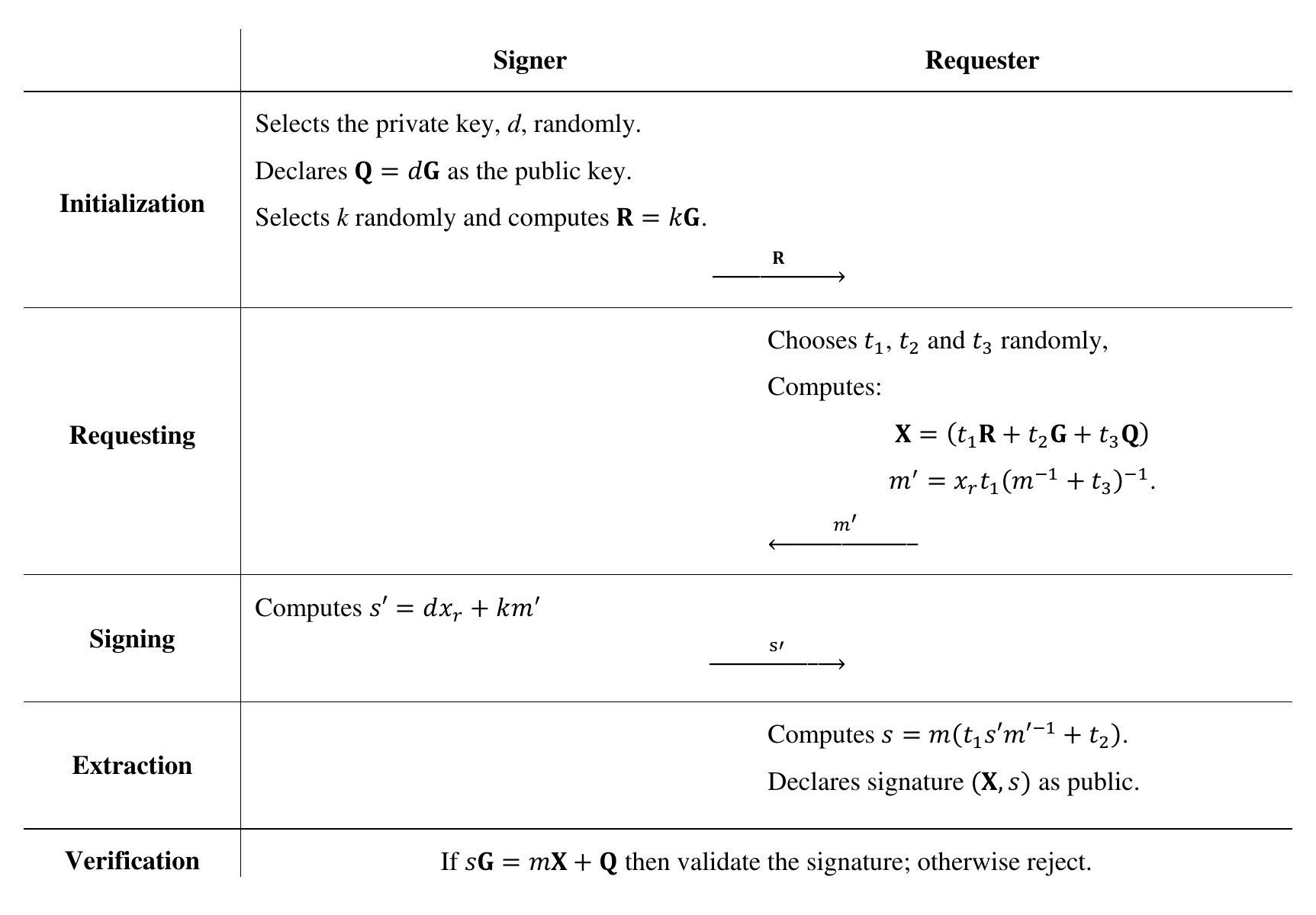}%
\caption{The proposed blind signature scheme.}%
\label{fig:scheme}%
\end{figure}

The correctness can be easily proved as follows:
\begin{align}
s\textbf{G}&=m(t_1 s'm'^{-1}+t_2 )\textbf{G} 		\nonumber \\
     &=m(t_1 (dx_r+km') x_r^{-1} t_1^{-1} (m^{-1}+t_3 )+t_2 )\textbf{G} \nonumber \\
     &=m(t_1 k+t_2+t_3 d)\textbf{G}+d\textbf{G} \nonumber \\
     &=m\textbf{X}+\textbf{Q}.
\end{align}

\subsection{Security of the Proposed Scheme}
The security of blind signature schemes is defined by unforgeability and blindness. Here, we discuss these properties of the proposed blind signature scheme.

\textbf{Unforgeability:} Forgery is an attack trying to fabricate a digital signature for a message without having access to the respective signer's private key. The security requirement of unforgeability of digital signatures is also called non-repudiation.

To forge a valid blind signature, the adversary should obtain the signature $s'$ or the signer's private key $d$ to fabricate the signature $s'=dx_r+km'$. It is impossible to obtain $d$ from the public key $\textbf{Q}$ using the equation $\textbf{Q}=d\textbf{G}$, because it is based on ECDLP. To forge $s'$, a dishonest requester (as an adversary) must calculate $dx_r+km'$. The requester knows the parameters $\textbf{Q}$ and $\textbf{R}$ and can compute $x_r \textbf{Q}+m'\textbf{R}$, which is equal to $s'\textbf{G}$. Again finding $s'$ from $s'\textbf{G}$ is impossible, because it is based on ECDLP. Thus, the unforgeability of the scheme is assured.

\textbf{Blindness:} A signature scheme is called blind, if the signer's view and the resulting signature are statistically independent. The signer's view is the set of all values that the signer gets during the execution of the signature issuing protocol, which in the proposed scheme is the tuple $(\textbf{R},m',s')$.

The three blinding functions are:
\begin{align}
\textbf{X}  &= (t_1 \textbf{R}+t_2 \textbf{G}+t_3 \textbf{Q}) \nonumber \\
m' &= x_r t_1 (m^{-1}+t_3 )^{-1}  \nonumber \\
s &=m(t_1 s'm'^{-1}+t_2 )
\end{align}

It can be seen that, there always exists a tuple of random numbers $(t_1,t_2,t_3)$ which maps any $(\textbf{R},m',s')$ to any $(\textbf{X},s)$, because there are three random parameters in the three blinding functions. Thus, the scheme is blind.

\subsection{Educed Schemes}
As in \cite{Huang}, three educed schemes are derived from the generalized scheme. In fact, two random parameters are sufficient to provide blindness. The tuple of random parameters $(t_1,t_2,t_3)$ for the three educed schemes are $(1,t_2,t_3)$, $(t_1,0,t_3)$ and $(t_1,t_2,0)$. The security of the educed schemes is discussed below.

\begin{itemize}
	\item Case I: $\mathbf{t_1=1}$
	
	In this case,the blinding functions are:
	\begin{align}
		\textbf{X} &= (\textbf{R}+t_2 \textbf{G}+t_3 \textbf{Q}) \nonumber \\
		m'&= x_r (m^{-1}+t_3 )^{-1} \nonumber \\
		s &=m(s'm'^{-1}+t_2 )
	\end{align}
	
	The correctness and the unforgeability are the same as the generalized scheme and the blindness can be proved as follows.
	
	Let $(x_r,m',s')$ be the data appearing in the signer's view during the execution of the signature and $(\textbf{X},s,m)$ be the corresponding data at the verifier. It is sufficient to show that there exist a pair of random numbers $(t_2,t_3)$ that maps $(x_{r_i},m'_i,s'_i )$ to $(\textbf{X}_j,s_j,m_j )$, for $i,j\in\{0,1\}$. We define:
	\begin{align}
		t_2 &= {m_j}^{-1}s_j-{m'_i}^{-1}s'_i \nonumber \\
		t_3 &= {m'_i}^{-1}x_{r_i}-{m_j}^{-1}
	\end{align}
By using Equations 6, 8, 10 and 14, we have:
\begin{align}
\textbf{R}_i+t_2 \textbf{G}+t_3 \textbf{Q} &= \textbf{R}_i+({m_j}^{-1} s_j-{m'_i}^{-1}s'_i)\textbf{G}+({m'_i}^{-1}x_{r_i}-{m_j}^{-1} )\textbf{Q} \nonumber \\
	&=\textbf{R}_i+{m_j}^{-1}s_j\textbf{G}-{m'_i}^{-1} (s'_i-dx_{r_i} )\textbf{G}-{m_j}^{-1} \textbf{Q} \nonumber \\
	&=\textbf{R}_i+{m_j}^{-1} (m_j \textbf{X}_j+\textbf{Q})-{m'_i}^{-1} (k_i m'_i )\textbf{G}-{m_j}^{-1} \textbf{Q} \nonumber \\
	&=\textbf{R}_i+\textbf{X}_j-k_i \textbf{G} \nonumber \\
	&=\textbf{X}_j 
\end{align}
Thus, the tuples $(x_{r_i},m'_i,s'_i )$ and $(\textbf{X}_j,s_j,m_j )$ have exactly the same relation defined by the signature issuing protocol, thus the scheme is blind.

	\item Case II: $\mathbf{t_2=0}$
	
	In this case, the blinding functions are:
	\begin{align}
		\textbf{X} &=(t_1 \textbf{R}+t_3 \textbf{Q}) \nonumber \\
		m' &=x_r t_1 (m^{-1}+t_3 )^{-1} \nonumber \\
		s &=t_1ms'm'^{-1}
	\end{align}
	
	The correctness and the unforgeability are also the same as the generalized scheme and the blindness is proved similar to the case I, by defining:
	\begin{align}
		t_1 &= {s'_i}^{-1} m'_i s_j m_j^{-1} \nonumber \\
		t_3 &= m_j^{-1} (x_{r_i} {s'_i}^{-1}s_j-1)
	\end{align}
	
	\item Case III: $\mathbf{t_3=0}$
	
	In this case, the blinding functions are:
	\begin{align}
		\textbf{X} &= (t_1 \textbf{R}+t_2 \textbf{G}) \nonumber \\
		m' &= x_r t_1 m \nonumber \\
		s &=m(t_1 s'm'^{-1}+t_2 )							
	\end{align}
	
	The correctness and the unforgeability are also the same as the generalized scheme and the blindness is proved similar to the case I, by defining: 
	\begin{align}
			t_1=m'_i {x_{r_i}}^{-1} {m_j}^{-1} \nonumber \\
			t_2=m_j^{-1} (s_j-{x_{r_i}}^{-1} s'_i )			
	\end{align}
\end{itemize}

\subsection{Performance of the Proposed Schemes}
The time complexity of the proposed schemes is compared with a recently proposed ECDLP-based blind signature \cite{Nikooghadam} and the scheme proposed by Camenisch et al. \cite{Camenisch}, which is declared to have superior performance than other DLP-based blind signatures \cite{Ting}.

Table \ref{tb:notations} defines the notations. In this table, the sub-index ($p$) denotes a prime field of order $2^p$. The time complexity of various operation units in terms of the time complexity of a modular multiplication is illustrated in Table \ref{tb:conversion} \cite{Chung}. Comparisons are based on the fact that an elliptic curve $E(F_q)$ with a point $\textbf{P}\in E(F_q)$ whose order is a 160-bit prime offers approximately the same level of security as DSA with a 1024-bit modulus $p$ \cite{Koblitz2}.

\begin{table}[!t]
\caption{Definition of Notations}
\label{tb:notations}
\centering
\begin{tabular}{|c|c|}
\hline
\bfseries \textbf{Notation} & \textbf{Definition}\\
\hline
$T_{\texttt{MUL}(p)}$ & Time complexity of a multiplication \\
\hline
$T_{\texttt{ADD}(p)}$ & Time complexity of an addition \\
\hline
$T_{\texttt{EXP}(p)}$ & Time complexity of an exponentiation \\
\hline
$T_{\texttt{inv}(p)}$ & Time complexity of an inversion \\
\hline
$T_{\texttt{EC-MUL}(p)}$ & Time complexity of an elliptic curve scalar multiplication \\
\hline
$T_{\texttt{EC-ADD}(p)}$ & Time complexity of an elliptic curve points addition \\
\hline
\end{tabular}
\end{table}

\begin{table}[!t]
\caption{Unit Conversion of Various Operations in Terms of $T_{\texttt{MUL}(1024)}$}
\label{tb:conversion}
\centering
\begin{tabular}{|c|c|}
\hline
\bfseries \textbf{Time Complexity of an Operation Unit} & \textbf{Time Complexity in Terms of Multiplication}\\
\hline
$T_{\texttt{EXP}(1024)}$	& $240\times T_{\texttt{MUL}}(1024)$  \\
\hline
$T_{\texttt{ADD}(1024)}$	& Negligible \\
\hline
$T_{\texttt{INV}(1024)}$	& $3\times T_{\texttt{MUL}(1024)}$  \\
\hline
$T_{\texttt{EC-MUL}(160)}$ & $29.3\times T_{\texttt{MUL}(1024)}$ \\
\hline
$T_{\texttt{EC-ADD}(160)}$ & $0.12\times T_{\texttt{MUL}(1024)}$ \\
\hline
$T_{\texttt{MUL}(160)}$ & $0.024\times T_{\texttt{MUL}(1024)}$ \\
\hline
$T_{\texttt{ADD}(160)}$ &	Negligible \\
\hline
$T_{\texttt{INV}(160)}$ & $0.073\times T_{\texttt{MUL}(1024)}$  \\
\hline
\end{tabular}
\end{table}

The detailed costs of the schemes are as follows:
\begin{align}
T_{\mbox{Camenisch \cite{Camenisch}}}		 &= 7T_{EXP(1024)}  +2T_{INV(1024)}  +10T_{MUL(1024)}+2 T_{ADD(1024)}				\nonumber \\
T_{\mbox{ECDLP-Based \cite{Nikooghadam}}}              &= 7T_{EC-MUL(160)}+3T_{EC-ADD(160)}+T_{INV(160)}   +6 T_{MUL(160)}+3 T_{ADD(160)}	\nonumber \\
T_{\mbox{Proposed}}							 				 &= 7T_{EC-MUL(160)}+3T_{EC-ADD(160)}+3 T_{INV(160)} +7 T_{MUL(160)}+3 T_{ADD(160)}	\nonumber \\	
T_{\mbox{Educed I}}				 							 &= 6T_{EC-MU(160)} +3T_{EC-ADD(160)}+3 T_{INV(160)} +5 T_{MUL(160)}+3 T_{ADD(160)} \nonumber \\
T_{\mbox{Educed II}}										 &= 6T_{EC-MUL(160)}+2T_{EC-ADD(160)}+3 T_{INV(160)} +7 T_{MUL(160)}+2 T_{ADD(160)}	\nonumber \\
T_{\mbox{Educed III}}			  						 &= 6T_{EC-MUL(160)}+2T_{EC-ADD(160)}+T_{INV(160)}   +7 T_{MUL(160)}+2 T_{ADD(160)}	
\end{align}

Table 3 provides a rough estimation of the overall time complexity of different schemes in terms of the required execution time for a modular multiplication. While maintaining the security, the proposed scheme is more efficient as compared to the scheme proposed by Camenisch et al. \cite{Camenisch} and has the same complexity as the ECDLP-based scheme proposed in \cite{Nikooghadam}. Also, the educed schemes are about 15\% more efficient than the generalized one.

\begin{table}[!t]
\caption{Time Complexity of Different Schemes in Unit of $T_{\texttt{MUL}(1024)}$}
\label{tb:complexity}
\centering
\begin{tabular}{|c|c|}
\hline
\bfseries \textbf{Scheme} & \textbf{Rough Estimation of the Computation Cost}\\
\hline
Camenisch \cite{Camenisch} 	& $1696\times T_{\texttt{MUL}(1024)}$\\
\hline
ECDLP-based \cite{Nikooghadam} & $206\times T_{\texttt{MUL}(1024)}$ \\
\hline
The proposed scheme & $206\times T_{\texttt{MUL}(1024)}$ \\
\hline
Educed scheme, Case I & $176\times T_{\texttt{MUL}(1024)}$ \\
\hline
Educed scheme, Case II & $176\times T_{\texttt{MUL}(1024)}$ \\
\hline
Educed scheme, Case III & $176\times T_{\texttt{MUL}(1024)}$ \\
\hline
\end{tabular}
\end{table}

\section{Conclusion}
This paper suggested a secure and efficient GOST-like blind signature scheme and three practical educed schemes based on the Elliptic Curve Discrete Logarithm Problem. The schemes utilize the inherent advantage of Elliptic Curve Cryptosystems in terms of smaller key size and lower computational overhead compared to its counterpart public key cryptosystems such as RSA and ElGamal. We proved the security of the proposed schemes is based on ECDLP and the time complexity is lower than the existing blind signature schemes. The schemes are applicable in the cryptographic services that emphasize the privacy of users, such as electronic voting over internet and untraceable payment services.

\end{document}